\documentstyle[aps,prl,epsfig,psfig,floats]{revtex}
\begin{document}
\draft
\twocolumn[\hsize\textwidth\columnwidth\hsize
\csname @twocolumnfalse\endcsname
\title{Theory for Electron-Doped Cuprate Superconductors:\\
{\it d}-wave symmetry order parameter}
\author{D. Manske$^1$, I. Eremin$^{2}$, and K.H. Bennemann$^1$}
\address{$^1$Instit\"ut f\"ur Theoretische Physik,
Freie Universit\"at Berlin,
Arnimallee 14, D-14195 Berlin, Germany}
\address{$^2$ Physics Department, Kazan State University,
Kremlyovskaya 18, 420008 Kazan, Russia}
\date{\today}
\maketitle
\begin{abstract}
Using as a model the Hubbard Hamiltonian we determine various 
basic properties of electron-doped cuprate superconductors like 
$\mbox{Nd}_{2-x}\mbox{Ce}_{x}\mbox{CuO}_{4}$ and 
$\mbox{Pr}_{2-x}\mbox{Ce}_{x}\mbox{CuO}_{4}$ for a
spin-fluctuation-induced
pairing mechanism. Most importantly 
we find a narrow range of superconductivity and
like for hole-doped cuprates $d_{x^{2}-y^{2}}$ - symmetry for 
the superconducting order parameter. The superconducting 
transition temperatures  $T_{c}(x)$ for various electron doping 
concentrations $x$ are calculated to be 
 much smaller than for hole-doped cuprates due to the different
Fermi surface and a flat band well below the Fermi level.
Lattice disorder may sensitively distort the symmetry 
$d_{x^{2}-y^{2}}$ via electron-phonon interaction. 

\end{abstract}
]
\narrowtext
One expects on general physical grounds if Cooper-pairing is
controlled  by antiferromagnetism that $d$-wave symmetry
pairing should also occur for electron-doped cuprates
\cite{remark1}. Until recently
\cite{tsuei,kokales,prozorov} experiment 
did not clearly support this and reported mainly $s$-wave
pairing \cite{hackl,tunnel,anlage}. Maybe as
a result of this, so far electron-doped cuprates received much less
attention than  hole-doped cuprates. Previously, we were rather
successful in determining the doping dependence of antiferromagnetism
in both electron- and hole-doped 
cuprates by using the Hubbard Hamiltonian \cite{baum}.
Applying this model to the hole-doped cuprates, many physical
quantities like the normal-state pseudogap and the doping
dependence of $T_c$ can also be described \cite{dmt,manske}.

Hence, to get an uniform theory we use here for 
the superconducting properties of  electron-doped 
cuprates also as a model the
2D one-band Hubbard Hamiltonian
\begin{equation}
H = - \sum_{\langle ij \rangle \, \sigma}
t_{ij}\left( c_{i\sigma}^+ c_{j\sigma} +
c_{j\sigma}^+ c_{i\sigma}\right)
+ U\, \sum_i n_{i\uparrow}n_{i\downarrow}
\quad .
\label{eq:hubbard}
\end{equation}
Here, $c_{i\sigma}^+$ creates an electron with spin $\sigma$
on site $i$, $U$ denotes the on-site Coulomb interaction,
and $t_{ij}$ is the hopping integral.
For the optimally doped NCCO the dispersion $\epsilon_{k}$ 
and Fermi surface are taken in accordance with 
photoemission (ARPES) experiments \cite{king}. Thus, we 
choose the
parameters $t=138$ meV and $t'=0.30$ in calculating
\begin{equation}
\epsilon_k = -2t \left[\cos k_x + \cos k_y - 2t'
\cos k_x \cos k_y + \mu/2 \right]
\quad ,
\label{eq:dispersion}
\end{equation}
where the chemical potential $\mu$ describes the band filling.
Here and in the following, we set the lattice constant $a=b$
equal to unity. 

\begin{figure}[t]
\vspace{-0.5cm}
\centerline{\epsfig{clip=,file=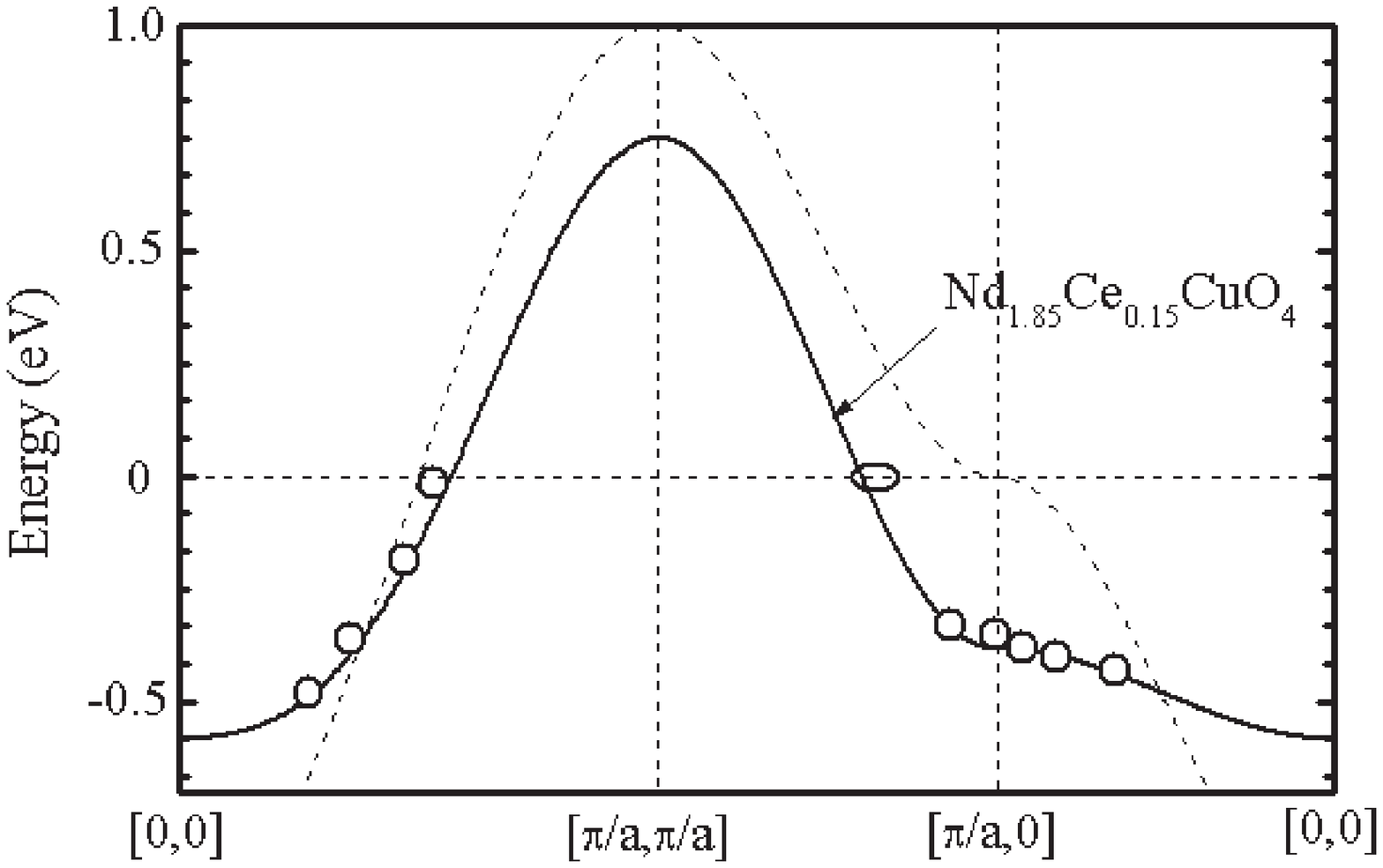,width=8.7cm,angle=0}}
\vspace{1ex}
\caption{Comparison of the energy dispersion $\epsilon_{k}$
for electron-doped cuprates and for hole-doped ones. 
Data (open dots) are taken from Ref. \protect\onlinecite{king}.
The solid curve refers to our tight-binding calculation as
described in the text. The dashed curve corresponds to
Eq. (\protect\ref{eq:dispersion}) with $t=250$ meV and $t'=0$.}
\label{fig1}
\end{figure}

In Fig. \ref{fig1} the results for $\epsilon_{k}$ are shown. 
For comparison, the results of a  tight-binding calculation
with $t=250$ meV and $t'=0$, which is often used to describe
the hole-doped superconductors, is also displayed. One
immediately sees one important difference: in the case of NCCO
the flat band is approximately $300$ meV {\it below} the Fermi
level, whereas for the hole-doped case the flat band lies very
close to it. Thus, one expects a smaller $T_c$ for electron-doped
cuprates than for the hole-doped cuprates.
Then, using $\epsilon_{k}$ in a spin-fluctuation-induced pairing
theory in the framework of the so-called  FLEX approximation 
\cite{bulut,tewordt,bennemann}, we calculate
the doping dependence  $T_{c}(x)$ and some other basic properties. 

\begin{figure}
\centerline{\epsfig{clip=,file=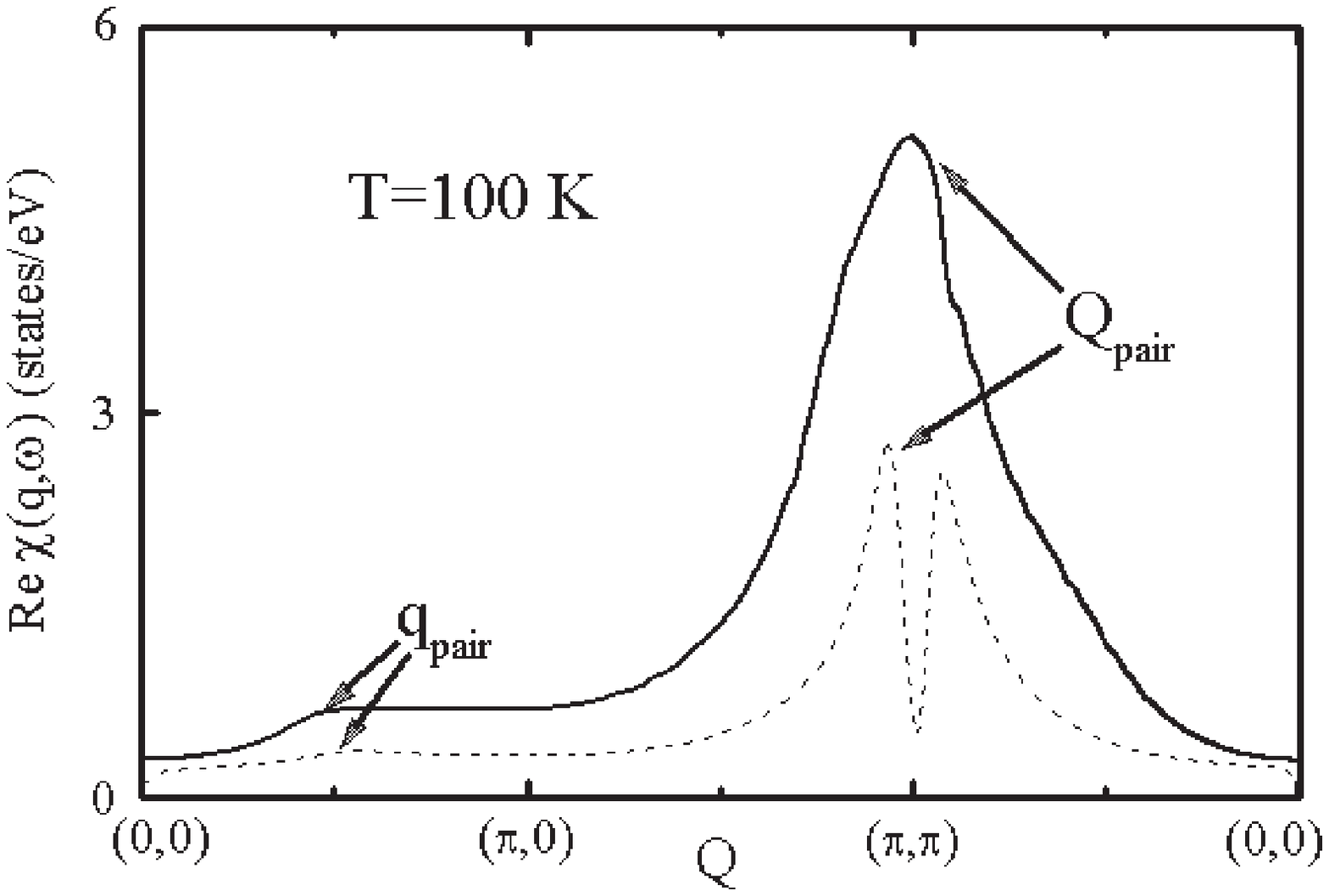,width=8.7cm,angle=0}}
\vspace{1ex}
\caption{Momentum dependence of the real part of the spin
susceptibility along the BZ route
$(0,0)\rightarrow(\pi,0)\rightarrow(\pi,\pi)\rightarrow(0,0)$
at T = 100 K for $\omega=0$ (solid curve) and $\omega =
\omega_{sf}\approx 0.47t$ (dashed curve). The main contributions 
to the corresponding pairing interaction come from ${\bf q}_{pair}$
(along the anti-nodes) and ${\bf Q}_{pair}$ (along the 'hot spots') 
as is illustrated in Fig. \protect\ref{fig5}.}
\label{fig2}
\end{figure}

In Fig. \ref{fig2} we show results for the real part of the
spin susceptibility at $100$K in the weak-coupling limit
for $\omega=0$ (solid curve) and for $\omega\ = \omega_{sf}
\approx 0.47t$ (dashed curve). $\omega_{sf}$ denotes the spin
fluctuation (paramagnon) energy, where a peak in $\mbox{Im }
\chi({\bf Q},\omega)$ occurs. The commensurate structure of 
$Re \chi({\bf q}, \omega = 0)$ is in accordance with recent 
calculations in Ref. \cite{ueda}, where it was pointed out
that the exchange of spin fluctuations yield a good description
of the normal state Hall coefficient R$_{H}$ for both hole- and
electron-doped cuprates. Furthermore, we also find a linear 
temperature dependence of the in-plane resistivity $\rho_{ab}(T)$,
if we do not take into account an additional electron-phonon
coupling. This will be discussed later. Concerning the 
superconducting properties, it was stated in 
Ref. \onlinecite{eremin}
that in contrast to the hole-doped superconductors
the electron-doped systems may be also close to a $d_{xy}$-symmetry
instability. However, within the picture of a
spin-fluctuation-induced pairing this is definitely not the case.
Since the  lower tiny peak favors $d_{xy}$ pairing symmetry and
the dominating larger peak $d_{x^{2}-y^{2}}$ symmetry (but
is pair-breaking for $d_{xy}$-symmetry), one
understands why an underlying superconducting order 
parameter $\phi( {\bf k}, \omega)$ exhibits almost pure
$d_{x^{2}-y^{2}}$ symmetry.

\begin{figure}
\centerline{\epsfig{clip=,file=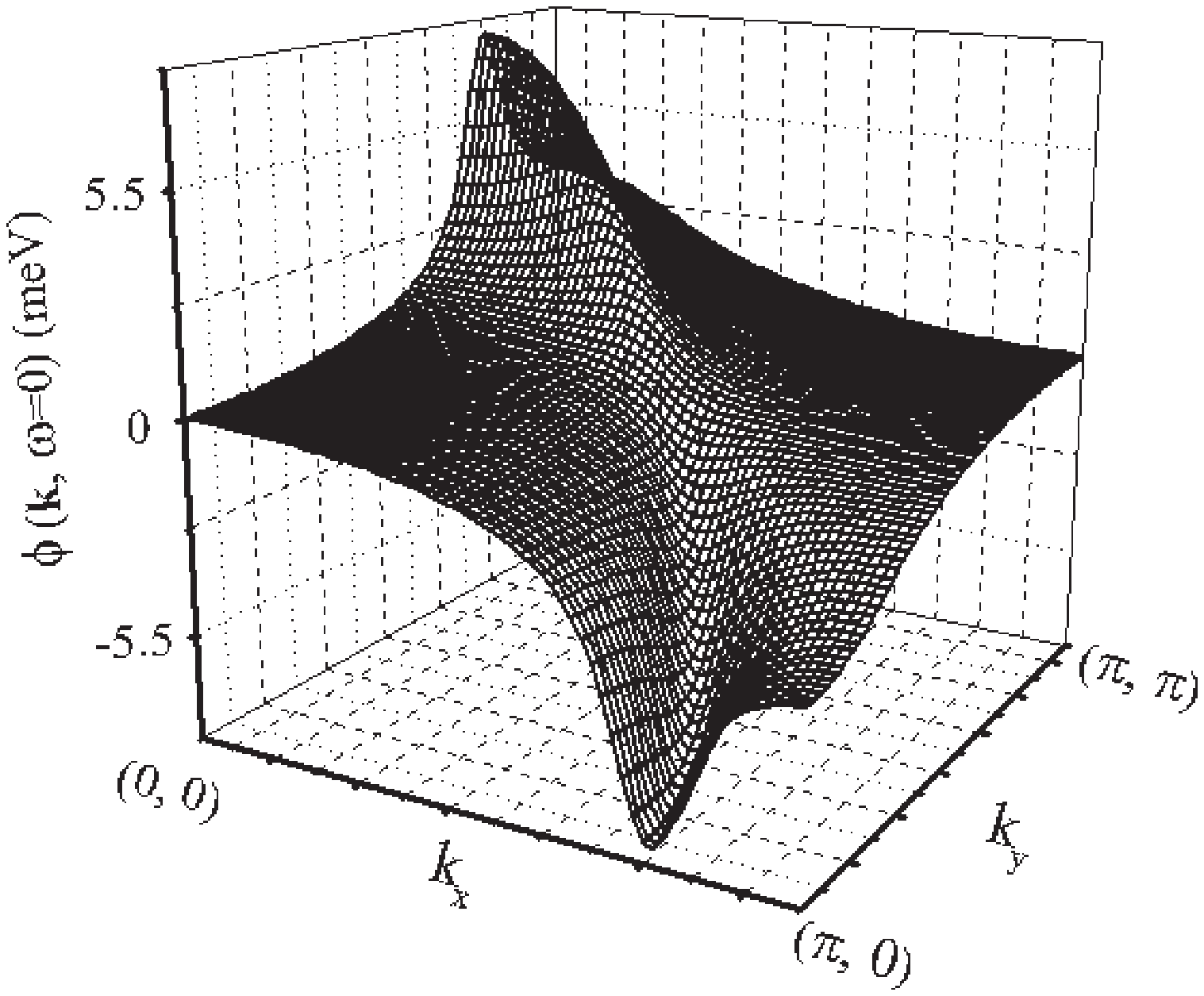,width=8.7cm,angle=0}}
\vspace{1ex}
\caption{ Calculated $d_{x^{2}-y^{2}}$-wave symmetry of the 
superconducting order parameter at T/T$_{c}$=0.8 for x=0.15 in the
first square of the Brillouin Zone. }
\label{fig3}
\end{figure}
In Fig. \ref{fig3} we present our result for $\phi({\bf k}, \omega)$ 
for $\omega=0$ and a doping $x=0.15$
at $T/T_c=0.8$, where the gap has just opened. The gap function
has clearly $d_{x^{2}-y^{2}}$-wave symmetry. This is in agreement
with the reported linear dependence of the in-plane
penetration depth for low temperatures \cite{kokales,prozorov}
and with phase-sensitive measurements \cite{tsuei}. From our obtained
result of a pure $d_{x^{2}-y^{2}}$-wave superconducting order
parameter we expect a zero-bias conductance peak (ZBCP) \cite{reiner} as 
observed for the hole-doped superconductors \cite{tunnel}. However,
its absence in the electron-doped cuprates may be attributed
to small changes in the surface quality and roughness \cite{apelbaum} 
or to disorder \cite{aprili}.
Note, the incommensurate structure in the order parameter close
to $(\pi,0)$ results from the double peak structure in
$\mbox{Re }\chi$ at $\omega \approx \omega_{sf} = 0.47t$
shown in Fig. \ref{fig2}. Physically, it means that Cooper-pairing
occurs not only for ${\bf Q}=(\pi,\pi)$, but mostly for
$\omega=\omega_{sf}$ and for
${\bf Q}^{*}=(\pi-\delta,\pi+\delta)$. Furthermore,
from Fig. \ref{fig2}, Fig. \ref{fig3},
and Fig. \ref{fig5} we conclude that {\it no} $d_{xy}$-symmetry
\begin{figure}[t]
\centerline{\epsfig{clip=,file=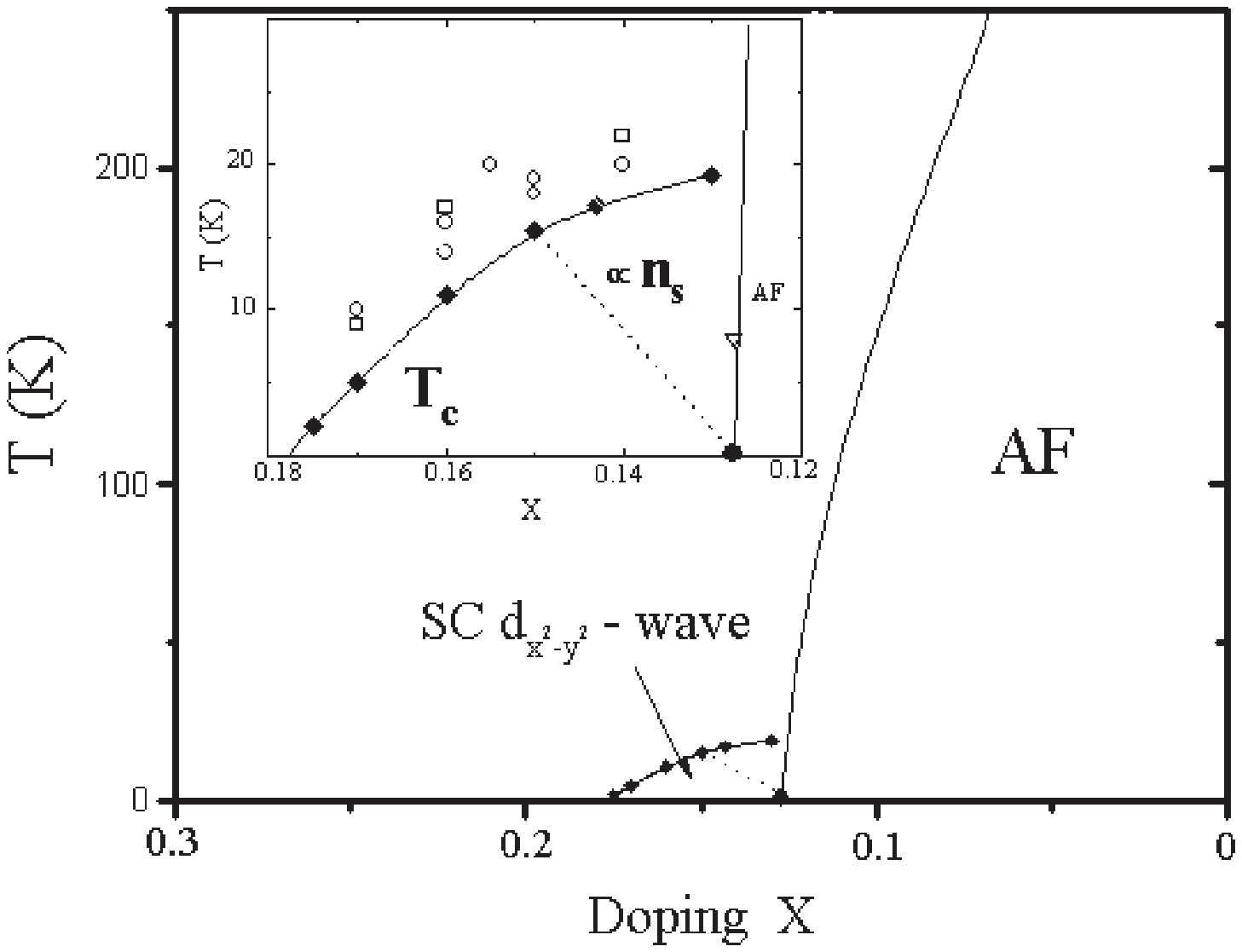,width=8.7cm,angle=0}}
\vspace{1ex}
\caption{Phase diagram $T(x)$ for electron-doped cuprates.
The AF transition line is taken from Ref.
\protect\onlinecite{baum}. Inset: blow-up of the
doping region $0.18<x<0.12$. The solid curve corresponds to
our calculated  T$_{c}$ values obtained from $\phi({\bf k}, \omega)=0$. 
For a comparison, also
experimental data are shown (squares
from Ref. \protect\onlinecite{paulus}, circles from Ref.
\protect\onlinecite{takagi}, triangle from
\protect\onlinecite{liang}). The dotted curve
refers to $T_{s}\propto n_{s}$.}
\label{fig4}
\end{figure}
component is present in the superconducting order parameter,
since the dominating $d_{x^{2}-y^{2}}$-type pairing suppresses
$d_{xy}$ pairing. ARPES study might test this.

In Fig. \ref{fig4} we present our results for the phase diagram
$T_c(x)$ and $T_N(x)$. We find that in comparison
to hole-doped superconductors smaller $T_c$ values and
superconductivity in a narrower doping range as is also observed in 
experiment \cite{almasan}.
Responsible for this are poorer nesting properties
of the Fermi surface and a flat band around $(\pi,0)$ which lies
well below the Fermi level. The narrow doping range is due to
antiferromagnetism up to $x=0.13$ and, for increasing $x$, rapidly
decreasing nesting properties. We have calculated the Cooper-pair
coherence length $\xi_{0}$, i.e. the size of a Cooper-pair, and
find similar values for electron-doped and hole-doped
superconductors (from 6 {\AA} to 9 {\AA}). If due to strong coupling lifetime
effects the superfluid density $n_{s}$ becomes small, the
distance $d$ between Cooper pairs increases.
If for $0.15 > x >0.13$ the
Cooper-pairs do not overlap significantly, i.e. $d/\xi_{0}>1$, 
then Cooper-pair
phase fluctuations get important \cite{chakraverty,emery,manske}.
Thus we expect like for hole-doped superconductors 
$T_c \propto n_s$. Assuming that $n_s$ increases approximately 
linearly from x$\simeq$0.13 to x$\simeq$0.15
we estimate a T$_{c}$ which is smaller than calculated from
$\phi({\bf k}, \omega)=0$ (see Fig. 4). As a consequence more experiments 
determining T$_{c}$ for x$\leq$0.15 should be performed
to check on the Uemura scaling $T_c \propto n_s$.

The effect of electron-lattice coupling on superconductivity
should depend on lattice perturbations like oxygen deficiencies.
Then, the isotope effect may show a distinct effect of
electron-phonon coupling on $T_{c}$.
On general grounds we expect a weakening of the
$d_{x^2-y^2}$-pairing
symmetry if we include the electron-phonon interaction and if this
plays a significant role. The absence of an isotope effect 
($\alpha_0=d\ln{T_c}/d\ln{M}\approx 0.05$) for
doping $x=0.15$ (see Ref. \onlinecite{batlogg}) suggests 
the presence of a pure $d_{x^2-y^2}$-symmetry.
We know from Fig. \ref{fig2} that phonons connecting the Fermi
surface with wave vector ${\bf Q}_{\mbox{pair}}=(\pi,\pi)$ will 
add destructively
to the spin fluctuation pairing \cite{dahm}. If, due to exchange of 
spin fluctuations, a $d_{x^{2}-y^{2}}$-symmetry instability is the
dominant contribution to the pairing interaction, an additional
electron-phonon coupling with wave vector 
${\bf q}_{\mbox{pair}}=(0.5\pi,0)$ would be also pair building. 
Note, we generally expect that due to the poorer nesting the
pairing instability due to electron-phonon and spin fluctuation
interaction become more easily comparable. In this case, the electron-
phonon coupling would definitely favor $s$-wave symmetry of the
underlying superconducting order parameter.
This can be analyzed
in detail by adding a term
$\alpha^{2}F(q,\omega)$ to the pairing interaction \cite{dahm}.
The corresponding phonon modes were calculated in Ref. \cite{heyen}.
Moreover, the inclusion of an electron-phonon interaction yields
a quadratic term in the resistivity for lower temperatures \cite{dahm}
as it is observed in experiment \cite{peng}.

To continue the discussion why the symmetry of the order parameter depends
for electron-doped cuprates more sensitively on electron-phonon
interaction, we show in Fig. \ref{fig5} the calculated Fermi surface 
for optimally doped NCCO.
Note, the topology of the Fermi surface for the electron-doped cuprates
is very similar to optimally hole-doped
Bi$_{2}$Sr$_{2}$CaCu$_{2}$O$_{8+\delta}$ (BI2212) as it was
also pointed out recently in Ref. \cite{maekawa}. 
We estimate that mainly no phonons are present along the edges
$(-0.25\pi, \pi) \rightarrow  (0.25\pi, \pi)$ 
bridging BZ areas, where the superconducting order parameter, 
$\phi({\bf k},\omega)$,
is always positive (denoted by $+/+$). Note, attractive
electron-phonon coupling bridging $+/-$ areas
$(-0,5\pi, -0.5\pi) \rightarrow (0.5\pi,0.5\pi)$ 
is {\it destructive} for 
$d_{x^{2}-y^{2}}$ Cooper pairing.
\begin{figure}[t]
\centerline{\epsfig{clip=,file=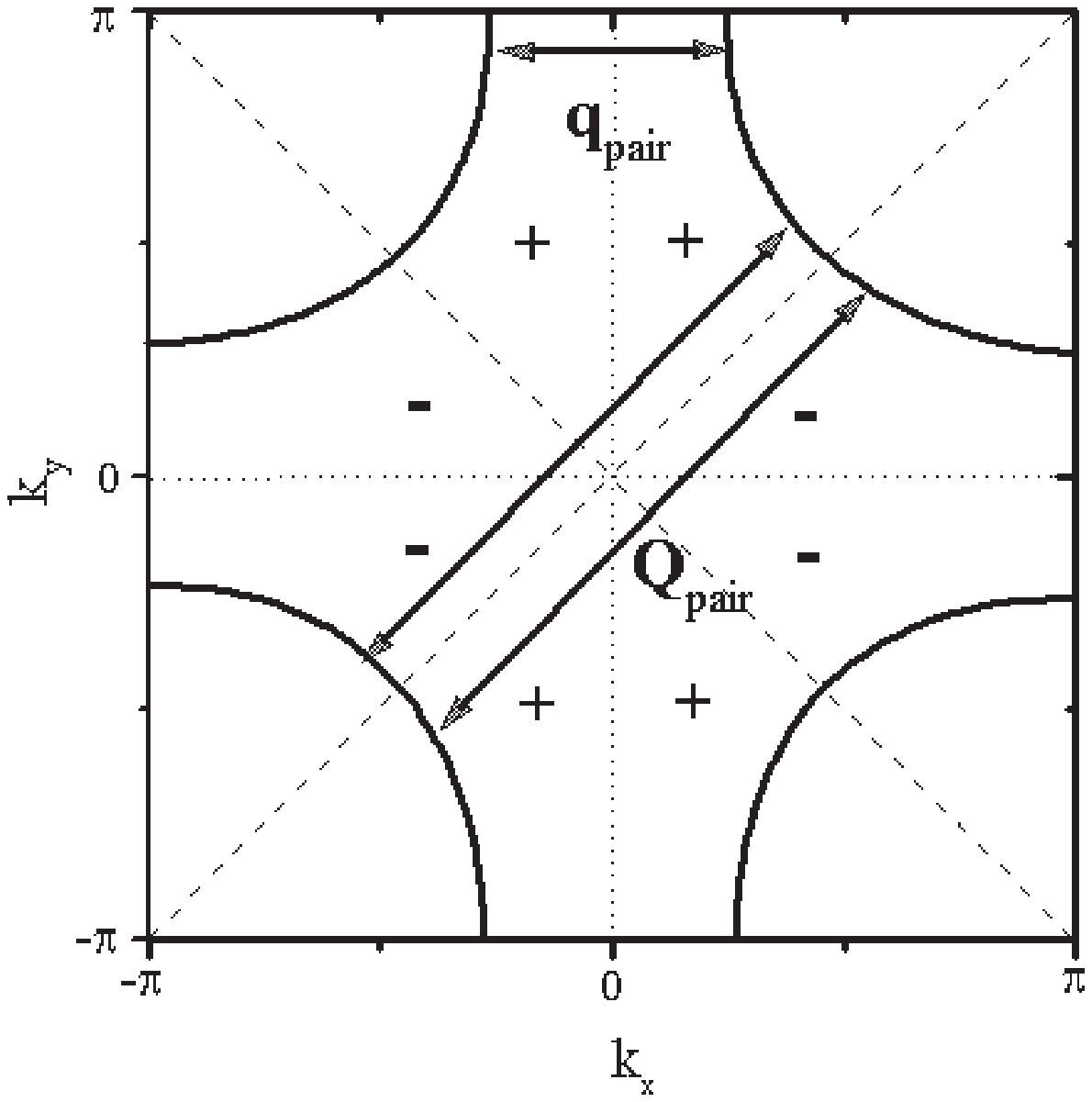,width=8.7cm,angle=0}}
\vspace{1ex}
\caption{Calculated Fermi surface for (optimally doped) 
$\mbox{NCCO}$. 
The $+(-)$ sign and the dashed curve corresponds to
the calculated momentum dependence (see Fig. \protect\ref{fig3})
of the the $d_{x^2-y^2}$ gap function $\phi(k,\omega=0)$ and its 
nodes, respectively.}
\label{fig5}
\end{figure}
However, due to poorer nesting conditions,
pairing transitions of the type $+/-$ are somewhat contributing
 and then a mixed symmetry \{$d_{x^{2}-y^{2}}+\alpha s$\} 
may occur.

Further experimental study of the doping dependence of the
oxygen-isotope effect are necessary for a better understanding
of the role played by the electron-phonon interaction. For example,
if due to
structural distortion and oxygen deficiency in the
$\mbox{CuO}_2$-plane the phonon spectrum $F(q,\omega)$ changes
significantly, then this affects $\alpha_0$ and reduces $T_c$.
Possibly the reported large isotope effect of $\alpha_0=0.15$ for
slightly changed oxygen content, 
i.e. $\mbox{Nd}_{1.85}\mbox{Ce}_{0.15}\mbox{CuO}_{3.8}$, 
could be related to this \cite{franck,onada}. As an example,
one might think of the oxygen out of plane B$_{2u}$ mode,
which become active if O$_{4}$ is replaced by O$_{3.8}$ 
\cite{heyen}.

In summary, our model for electron-doped cuprates yields like
for hole-doped case  pure $d_{x^{2}-y^{2}}$ symmetry pairing in 
a good agreement with recent experiments. In contrast to 
hole-doped superconductors, we find for electron-doped cuprates
smaller T$_{c}$ values due to a flat band dispersion around 
$(\pi,0)$ well below the Fermi level. Futhermore, superconductivity
only occurs for a narrow doping range $0.18>x>0.13$ because of the 
onset of antiferromagnetism, and, on the other side, due to poorer
nesting conditions. We get $2\Delta/k_{B}T_{c}=5.3$ for
$x=0.15$ in reasonable agreement with Ref. \cite{hackl}. We argue that
if the electron-phonon coupling becomes important, for example
due to oxygen deficiency, then the $s$-wave pairing instability
competes with $d_{x^{2}-y^{2}}$-wave symmetry. This might explain
a possible $s$-wave order parameter as reported in earlier measurements. 

Its pleasure to thank R. Hackl, M. Opel, L. Alff, K. Scharnberg, 
and T. Dahm for useful 
discussions. One of us (I. E.) would like to thank for the 
financial support
German Academic Exchange Service (DAAD), the Freie Universit\"at
Berlin  and Russian Scientific
Council on Superconductivity (Grant No. 98014).

\begin{references}
%
%
\bibitem{remark1} If the dominant {\it repulsive} pairing
contribution in high-$T_c$ superconductors can be mainly
described by their spin susceptibility, then the underlying
order parameter must change its sign. From group theory we
know \protect\cite{rice} that for a nested Fermi surface
described by ${\bf Q}=(\pi,\pi)$, i.e. $\epsilon_{\bf k+Q} =
-\epsilon_{\bf k}$, the $d_{x^2 - y^2}$-symmetry order
parameter is the simplest possibility.
%
\bibitem{rice} M. Sigrist and T.~M. Rice, Z. Phys. B -
Condensed Matter {\bf 68}, 9 (1987).
%
\bibitem{tsuei} C. C. Tsuei and J.~R. Kirtly,
cond-mat/0002341.
%
\bibitem{kokales} J. David Kokales {\it et al.},
cond-mat/0002300.
%
\bibitem{prozorov} R. Prozorov, R. W. Gianetta, P. Furnier, 
and R. L. Greene, cond-mat/0002301.
%
\bibitem{hackl} B. Stadlober {\it et al.}, Phys. Rev. Lett.
{\bf 74}, 4911 (1995).
%
\bibitem{tunnel} L. Alff {\it et al.}, Phys. Rev. B.
{\bf 58}, 11197 (1998).
%
%
\bibitem{anlage} S. M. Anlage {\it et al.}, Phys. Rev. B.
{\bf 50}, 523 (1994).
%
\bibitem{baum} G. Baumg\"artel, J. Schmalian, and 
K. H. Bennemann,
Phys. Rev. B {\bf 48}, 3983 (1993).
%
\bibitem{dmt} T. Dahm, D. Manske, and L. Tewordt,
Phys. Rev. B {\bf 55}, 15274 (1997).
%
\bibitem{manske} D. Manske, T. Dahm, and K.~H. Bennemann,
cond-mat/9912062.
%
\bibitem{king} D.~M. King {\it et al.}, Phys. Rev. Lett.
{\bf 70}, 3159 (1993).
%
%
\bibitem{bulut} N. E. Bickers, D. Scalapino, and S. R. White, 
Phys. Rev. Lett. {\bf 62}, 961 (1989).
%
%
\bibitem{tewordt} T. Dahm, and L. Tewordt, 
Phys. Rev. Lett. {\bf 74}, 793 (1995).
%
\bibitem{bennemann} M. Langer, J. Schmalian, S. Grabowski, and 
K.-H. Bennemann, Phys. Rev. Lett. {\bf 75}, 4508 (1995).
%
\bibitem{ueda} H. Kontoni, K. Kanki, and K. Ueda, Phys. Rev. B
{\bf 59}, 14723 (1999).
%
\bibitem{eremin} K. Kuroki and H. Aoki, J. Phys. Soc. Jpn. {\bf 67},
1533 (1998).
%
\bibitem{reiner} M. Fogelstrom, D. Reiner, and J. A. Sauls,
Phys. Rev. Lett. {\bf 79}, 281 (1997).
%
\bibitem{apelbaum} J. A. Appelbaum, Phys. Rev. {\bf 154}, 633 (1967).
%
\bibitem{aprili} M. Aprili, M. Covington, E. Paraoani, 
B. Niedermeier, and L. H. Greene, Phys. Rev. B {\bf 57}, 8139 (1998).
%
\bibitem{almasan} C. Almasan, and M. B. Maple, in {\it Chemistry
of High-Temperature Superconductors}, ed. by C. N. R. Rao (World
Scientific, Singapore), 1991.
%
\bibitem{paulus} E.~F. Paulus {\it et al.}, Solid State Comm.
{\bf 73}, 791 (1990).
%
\bibitem{chakraverty} B.~K. Chakraverty, A. Taraphder, and M.
Avignon, Physica C {\bf 235-240}, 2323 (1994).
%
\bibitem{emery} V.~J. Emery and S.~A. Kivelson, Nature
{\bf 374}, 434 (1995).
%
\bibitem{takagi} H. Takagi, S. Uchida, and Y. Tokura,
Phys. Rev. Lett. {\bf 62}, 1197 (1989).
%
\bibitem{liang} G. Liang {\it et al.}, Phys. Rev. B
{\bf 40}, 2646 (1989).
%
\bibitem{batlogg} B. Batlogg {\it et al.}, Physica C
{\bf 185-189}, 1385 (1991).
%
\bibitem{dahm} T. Dahm, D. Manske, D. Fay, and L. Tewordt,
Phys. Rev. B {\bf 54}, 12006 (1996).
%
\bibitem{heyen} E. T. Heyen {\it et al.}, Solid State Comm. 
{\bf 74}, 1299 (1990).
%
\bibitem{peng} J. L. Peng, E. Maiser, T. Venkatesan, 
R. L. Greene, and G. Czyzek, Phys. Rev. B {\bf 55}, 6145 (1997).
%
\bibitem{maekawa} T. Tohoyama and S. Maekawa, Supercond. Sci. 
Technol. {\bf 13}, R17 (2000).
%
\bibitem{franck} J. P. Franck, in {\it Physical properties 
of High Temperature Superconductors}, ed. D. Ginsberg, 
(World Scientific, Singapore), 1994.
%
\bibitem{onada} M. Onada, S. Kondoh, and M. Sato, Solid State Comm. 
{\bf 70}, 1141 (1989).
%
\end{references}
\end{document}